\documentstyle[epsfig]{article}
\def\parno {\par\noindent}
\begin{document}
\begin{centering}
%
%
{ \bf    SPECTRAL DUALITY AND DISTRIBUTION OF EXPONENTS\\ 
                   FOR TRANSFER MATRICES\\
              OF BLOCK TRIDIAGONAL HAMILTONIANS }
\centerline {(revised, february 2003)}
\vskip 0.5truecm
Luca Molinari\\
Dipartimento di Fisica dell'Universit\`a degli Studi di Milano\\
and I.N.F.N., sezione teorica di Milano. Via Celoria 16, I-20133 Milano\\
luca.molinari@mi.infn.it\\
\end{centering}
\vskip 0.5truecm\noindent
{\bf Abstract:}  

I consider a general block-tridiagonal matrix and the corresponding
transfer matrix. By allowing for a complex Bloch parameter in the boundary 
conditions, the two matrices are related by a spectral duality.\parno
As a consequence, I derive some analytic properties of the exponents of the 
transfer matrix in terms of the eigenvalues of the (non-Hermitian) block
matrix. Some of them are the single-matrix analogue of results 
holding for Lyapunov exponents of an ensemble of block matrices, that occur
in models of transport.\parno
The counting function of exponents is related to winding numbers of 
eigenvalues.  I discuss some implications of duality on the distribution 
(real bands and complex arcs) and the dynamics of eigenvalues.\parno
P.A.C.S.: 02.10.Yn (matrix theory), 72.15.Rn (localization effects), 
72.20.Ee (mobility edges, hopping transport)
\vskip 1truecm
{\bf 1. Introduction}\par
A tridiagonal Hermitian matrix whose entries are square matrices of size $M$ 
is a block tridiagonal matrix. By denoting $H_i=H_i^\dagger$ the diagonal 
blocks and by $L_i$ the blocks in the next upper diagonal, the eigenvalue 
equation written in the block components of an eigenvector $\underline u$ is:
\begin{eqnarray}
 H_i \underline u_i + L_i \underline u_{i+1} + L_{i-1}^\dagger 
\underline u_{i-1} = E\underline u_i \label{tb} 
\end{eqnarray}
We may view $H_i$ as Hamiltonian matrices of a chain of 
subsystems, each with $M$ internal states, sequentially coupled 
by matrices $L_i$. We shall make the only requirement $\det L_i\neq 0$.
\parno
Band matrices are in this class, with non-diagonal blocks being triangular.
Random band matrices are studied 
for quantum chaos and transport, mainly by numerical means \cite{Cas} or 
by a mapping on a non-linear supersymmetric sigma model \cite{Fyo}.
The block structure is typical of tight binding models, as Anderson's model 
for the transport of a particle in a lattice with impurities \cite{Kra}. Here,
the matrices $H_i$ are the Hamiltonians of isolated slices transverse to 
some direction and the matrices $L_i$ contain the hopping amplitudes between 
lattice sites of neighboring slices. Tridiagonal arrays of random matrices 
were studied in the context of multichannel scattering in mesoscopic or 
nuclear physics \cite{Wei} or as multi-matrix models in the large size 
limit \cite{Bre}. For recent reviews of applications of random matrices 
in physics, see \cite{Guhr,Bee}.\par
The second-order recursive character of (\ref{tb}) makes it useful to 
introduce a transfer matrix $T_i(E)$, of size $2M\times 2M$ 
\begin{eqnarray}
 \pmatrix {\underline u_{i+1} \cr \underline u_i \cr } = 
 \pmatrix { L_i^{-1}(E-H_i) & -L_i^{-1}L_{i-1}^\dagger\cr I & 0\cr } 
 \pmatrix {\underline u_i \cr \underline u_{i-1}\cr }  
\end{eqnarray}
The block components of the eigenvector are reconstructed by applying 
a product of transfer matrices to an initial block pair. 
A length $N$ of the chain corresponds to the transfer matrix 
$T(E)=T_N(E)\ldots T_1(E)$:
\begin{eqnarray}
 \pmatrix {\underline u_{N+1} \cr \underline u_N \cr} = T(E) 
\pmatrix {\underline u_1\cr \underline u_0 \cr } \label{trm}  
\end{eqnarray}
The transfer matrix is the main tool for investigating boundary
properties of the Hamiltonian's eigenvectors, or the transmission matrix 
of the chain coupled to leads. I developed a theory for transfer matrices 
of general block-tridiagonal Hamiltonians in \cite{Mol1,Mol2}. Recently the 
formalism was applied to transport in nanotubes and molecules and 
generalized to allow for non-invertible off-diagonal blocks 
\cite{Mah,Kost1,Kost2,HjS}.\parno
In this paper I describe some interesting consequences of a nice algebraic 
identity involving the characteristic polynomials of the two matrices 
\cite{Mol1}. Though this spectral duality holds in general, I here restrict 
to Hermitian block tridiagonal matrices because of their relevance in 
physics.\par
For a chain of length $N$ we must provide boundary conditions. With periodic
boundary conditions we require $L_0=L_N$ for Hermiticity and 
$\underline u_0=\underline u_N$, $\underline u_{N+1}=\underline u_1$.
However, it turns to be very convenient to allow for a complex Bloch 
parameter: 
\begin{eqnarray}
\underline u_0= {1\over z} \underline u_N, \qquad
\underline u_{N+1}=z\underline u_1 \label{bc}
\end{eqnarray}
Therefore the Hamiltonian matrix is block-tridiagonal with corners:
\begin{eqnarray}
 {\cal H}(z) =\pmatrix {
  H_1     & L_1     & {} & {}     & {} & {} & {1\over z} L_N^\dagger  \cr
 L_1^\dagger & H_2  & L_2 & {}    & {} & {} &        {}      \cr
 {}       &  L_2^\dagger & \ldots & \ldots & {} & {} & {}     \cr
 {}       &     {}    &  {}  & {} & \ldots  & \ldots  & L_{N-1}     \cr
 z L_N    &     {}    &  {}  & {} & {} & L_{N-1}^\dagger   &  H_N  \cr }
\end{eqnarray}
It is Hermitian only for $|z|=1$. Boundary conditions 
with $z=e^{i\varphi}$ arise when decomposing the eigen-problem (\ref{tb}) for
an infinite periodic chain of period $N$ in the eigen-spaces of 
the $N$-block translation operator, as well as in the topology of
a $N$-site ring with a magnetic flux through it. In general it is: 
\begin{eqnarray}
{\cal H}(z)^\dagger ={\cal H}(1/z^*)
\end{eqnarray}
Non Hermitian tridiagonal matrices ($M=1$), with $z=e^{Ng}$ and $g\ge 0$, 
were introduced by Hatano and Nelson \cite{HN} in a study on vortex depinning 
in superconductors which promoted a burst of research,
see for example \cite{Schn,FZ,Gol1,Mud}. Spectral properties were 
analyzed in greater detail by Goldsheid and Khoruzhenko \cite{Gol3} who 
proved that, for  $g>g_{cr}$, eigenvalues corresponding to extended states
begin to migrate in the complex plane and distribute along the level curve 
of the single Lyapunov exponent of the model: $\gamma (E)=g$. Two wings of 
real eigenvalues correspond to exponentially localized eigenvectors, that 
are insensitive to the boundary.\par
These features also appear in the more difficult case of  block matrices. 
For $|z|$ sufficiently greater or smaller than unity, the block matrix 
develops complex eigenvalues which are seen numerically to distribute 
along lines. The eigenvalues of an eptadiagonal matrix (M=3) with 
diagonal disorder and unit hopping amplitudes, are shown in Fig.2.\par
It is intuitive that 
there is a connection between boundary properties of eigenvectors, which
are described by the transfer matrix, and the response of energy eigenvalues 
to variations of boundary conditions: namely, the Landauer and Thouless 
approaches to transport. This intuition has a formulation in the spectral 
duality.\parno
After a short review of duality (section 2), I derive several analytic  
properties of the exponents of a single transfer matrix (section 3). In 
particular, I evaluate the counting function of exponents as the winding
number of trajectories of eigenvalues of the source block matrix.
In section 4, I use duality to describe qualitatively the distribution and 
dynamics of eigenvalues of Hermitian and non-Hermitian block-tridiagonal 
matrices.
\vskip 1truecm

\noindent {\bf 2. Spectral Duality}\parno 
In this section I review some basic properties of a transfer matrix $T(E)$ of 
a chain of length $N$,  with $L_0=L_N$ \cite{Mol1, Mol2}.\vskip 0.5truecm
$T(E)$ is a matrix polynomial in $E$ of degree $N$, with a non-zero 
determinant independent of $E$ and of diagonal blocks $H_i$:
\begin{eqnarray}
\det T(E)= \prod_{i=1}^N {{\det L_i^\dagger}\over {\det L_i}} \label{det}
\end{eqnarray}
The block structure of the Hamiltonian and the corresponding factorization 
of the transfer matrix, makes a typical property of transfer matrices apparent:
\vskip 0.5truecm\noindent
{\bf Proposition 1:} {\underbar {the symplectic property}}.
\begin{eqnarray} 
 T(E^*)^\dagger \Sigma_N T(E) = \Sigma_N, \qquad 
\Sigma_N =\pmatrix {0 & -L_N^\dagger \cr  L_N & 0 \cr } \label{symp}
\end{eqnarray}
\noindent
Proof: it is a consequence of $L_0=L_N$ and of the factorization of $T(E)$ 
into a product of matrices $T_k(E)$, whose inverse is:
$ T_k(E)^{-1}=\Sigma_{k-1}^{-1}T_k(E^*)^\dagger \Sigma_k $. $\bullet$
\vskip 0.5truecm\noindent
{\bf Corollary}: if $z$ is an eigenvalue of $T(E)$, then  $1/z^*$ is an 
eigenvalue of $T(E^*)$. \parno
For real $E$, both $z$ and $1/z^*$ are in the spectrum of $T(E)$. 
If $T(E^*)=T(E)^*$, then both $z$ and $1/z$ are in the spectrum of $T(E)$.
\parno
These statements summarize in the useful identity:
\begin{eqnarray}
 \det[T(E)-z] = z^{2M} \det T(0) \det [T(E^*) -{1/z^*}]^* \label{det2} 
\end{eqnarray}
\\
Let $\underline u$, with blocks $\underline u_1,\ldots , 
\underline u_N$, be an eigenvector of ${\cal H}(z)$ with eigenvalue $E$. 
Then, by eq. (\ref{trm}) and after imposing the boundary conditions 
(\ref{bc}):
\begin{eqnarray}
  \pmatrix {z \underline u_1\cr  \underline u_N\cr } = T(E) \pmatrix {
\underline u_1\cr 1/z \underline u_N\cr } 
\end{eqnarray}
which means that $z$ is an eigenvalue of $T(E)$ with an eigenvector of blocks
$z\underline u_1$ and $\underline u_N$. However the converse is true:
given an eigenvector of T(E) with eigenvalue $z$ one reconstructs via 
products of matrices $T_k(E)$ the whole
eigenvector of ${\cal H}(z)$ with eigenvalue $E$. Therefore 
$\det[E-{\cal H}(z)]=0$ if and only if $\det[T(E)-z]=0$; this duality among
eigenvalues is made precise as an identity among  characteristic 
polynomials \cite{Mol1}:
\vskip 0.5truecm\noindent
{\bf Proposition 2:} {\underbar {the spectral duality}}.
\begin{eqnarray}
 \det [E -{\cal H}(z)] = (-z)^{-M}\det [L_1 L_2\ldots L_N]
\det [T(E)-z]  \label{dual}
\end{eqnarray}
\noindent
Proof: we must show that $\det [T(E)-z]$ is a polynomial in $E$ of degree 
$NM$ with leading coefficient $(-z)^M\det [L_1\ldots L_N]^{-1}$. To this end, 
we first consider the leading terms in $E$ of both sides of eq.(\ref{det2}): 
the equality implies that the leading term of $\det [T(E)-z]$ is proportional 
to $z^M$. Next we derive the following leading block stucture of $T(E)$:
$$ T(E) \approx \pmatrix { E^N Q & -E^{N-1}QL_N^\dagger \cr 
E^{N-1}L_N Q & -E^{N-2}L_NQL_N^\dagger \cr }, \quad 
 Q=(L_1\ldots L_N)^{-1} $$
The leading term in $E$ of $\det [T(E)-zI]$ with the constraint of being
proportional to $z^M$, is provided by the diagonal factors $\det (E^N Q)$
and $\det (-zI)$. $\bullet$
\vskip 0.5truecm \noindent
{\bf Corollary}: If Im$E\neq 0$ then $T(E)$ has no eigenvalues on the unit 
circle.\parno
Proof: for $|z|=1$ the matrix ${\cal H}(z)$ is Hermitian, and for
${\rm Im}E\neq 0$ it is always $\det[E-{\cal H}(z)]\neq 0$. By duality, this
implies $\det[T(E)-z]\neq 0$. $\bullet$
\vskip 0.5truecm
\noindent
Notes: in \cite{Mol2} I provided a representation of $T(E)-z$ in terms of 
the corner blocks of the resolvent of ${\cal H}(z)$. I also state the spectral 
duality for the matrix $T^\dagger T$.\parno 
In the tridiagonal case ($M=1$), blocks are numbers. If 
$\lambda =L_1\ldots L_N$ the spectral duality simplifies to the known 
expression 
\begin{eqnarray}
\det[E-{\cal H}(z)]= \lambda \,{\rm tr}T(E) -\lambda z -\lambda^* {1\over z}
\end{eqnarray}
\vskip 1truecm

\noindent{\bf 3. The spectrum of exponents}\parno
Let us fix $E$ real or complex and denote as $z_a(E) =e^{\xi_a +i\varphi_a}$,
$a=1\ldots 2M$, the eigenvalues of $T(E)$. The real numbers $\xi_a(E)$ are the
{\bf exponents} of the transfer matrix. For real $E$ the symplectic property 
(\ref{symp}) assures that exponents come in pairs $\pm\xi_a (E)$.\parno
The property $|\det T(E)|=1$, see (\ref{det}), implies: 
\begin{eqnarray}
\sum_{a=1}^{2M} \xi_a(E) = \, 0 \label{exp}
\end{eqnarray}\noindent
When considering an ensemble of random Hamiltonian matrices with tridiagonal
block structure, one is often interested 
in the corresponding ensemble of transfer matrices. Being a product of $N$ 
random matrices, a transfer matrix develops exponents $\xi_a(E)$ that  
asymptotically
grow linearly in the length $N$ \cite{Cri}, with a coefficient
known as the {\bf Lyapunov exponent}:
\begin{eqnarray}
  \gamma_a (E) =\lim_{N\to\infty}{1\over N} \langle \xi_a(E) \rangle
\end{eqnarray}
For tridiagonal random matrices there is just one pair of opposite
Lyapunov exponents, that can be evaluated with the Herbert-Jones-Thouless 
formula \cite{Thou}, with the knowledge of the average density of 
eigenvalues:
$$ \gamma (E) = const + \int dE^\prime \rho(E^\prime ) \log |E-E^\prime | $$
The extension to a complex value $E$ is discussed in \cite{Derr}.\parno
For the Anderson model \cite{Mar} or Band Random Matrices \cite{Kot}
there are several numerical studies of Lyapunov spectra. In these cases of 
great physical interest, transfer matrices are derived from the Hamiltonians 
and the analytic approach is difficult.\par
It is interesting that an analytic formula relating the distribution of  
the exponents to the spectrum of the Hamiltonian is possible. Note that
the following statements are true for a single and general block-tridiagonal 
Hamiltonian.\parno
We shall deduce several results from
\vskip 0.5truecm
 \noindent
{\bf Proposition 3:} 
\begin{eqnarray}
 \int_0^{2\pi} {{d\varphi }\over {2\pi}} \log | \det [E-{\cal H}
(e^{\xi+i\varphi})]| = \sum_{i=1}^N \log |\det L_i| +
{1\over 2}\sum_{a=1}^{2M} |\xi-\xi_a (E)|  \label{th1}
\end{eqnarray}
Proof: The duality relation (\ref{dual}) gives:\parno 
\begin{eqnarray}
\log |\det [E-{\cal H}(e^{\xi+i\varphi})]| - \sum_i\log |\det L_i| =
\nonumber \\
= -M\xi +{1\over 2} \sum_{a=1}^{2M} 
\log |e^{\xi_a+i\varphi_a} - e^{\xi+i\varphi} |^2 \nonumber \\
=  {1\over 2} \sum_{a=1}^{2M} \{
\xi_a + \log [2\cosh (\xi_a-\xi)-2\cos (\varphi_a-
\varphi )] \} \nonumber \\
= {1\over 2} \sum_{a=1}^{2M} |\xi_a-\xi| -\sum_{\ell=1}^\infty 
\sum_{a=1}^{2M} {{\cos \ell (\varphi_a-\varphi )}\over \ell}
e^{-\ell |\xi_a -\xi |} \label{FE}
\end{eqnarray}
We used eq.(\ref{exp}) and the formula (see eq. 1.448.2 in \cite{GR}):
$$ \log [2\cosh x-2\cos y]=|x|-2\sum_{\ell =1}^\infty 
{{\cos \ell y}\over \ell } e^{-\ell |x|} $$
Eq. (\ref{FE}) is the Fourier expansion  of $\log |\det [E-{\cal H}
(e^{\xi+i\varphi})]|$, which is a periodic function of $\varphi $. 
The constant mode is just the proposition. $\bullet $
\vskip 0.5truecm \noindent
In the special case $\xi=0$, the matrix ${\cal H}(e^{i\varphi})$ is Hermitian, 
and (\ref{th1}) yields a formula for the {\bf sum of positive exponents} 
($E$ can be complex):
\parno
{\bf Proposition 4}:
\begin{eqnarray}
 \sum_{\xi_a >0} \xi_a(E) = -\sum_i \log|\det L_i| + \int_0^{2\pi} 
{{d\varphi}\over {2\pi}}\log |\det (E-{\cal H}(e^{i\varphi}) )| 
\label{Soui1}
\end{eqnarray}
By taking the derivative in $\xi $ of (\ref{th1}) we obtain the 
{\bf spectral counting function}, which counts the exponents less than
$\xi $, for any complex value $E$;
\begin{eqnarray}
{\cal N}(\xi_a(E) \le \xi) = 
\sum_{a=1}^{2M} \theta (\xi -\xi_a(E)) =\\
M+ {d\over {d\xi}} \int_0^{2\pi} {{d\varphi }\over {2\pi}} 
\log | \det [E-{\cal H} (e^{\xi+i\varphi})]| \label{cf}
\end{eqnarray}
We now write $|\det [E-{\cal H}(z)]|$ in terms of the eigenvalues 
$E_n (z)$ and their complex conjugate, which equal $E_n(1/z^*)$,
and evaluate the derivative: 
\begin{eqnarray}
{d\over {d\xi}} \log | \det [E-{\cal H} (e^{\xi+i\varphi})] |= \nonumber \\
=\, -{1\over 2}\sum_n \left (
{1\over {E-E_n(z)}}{{\partial E_n(z)}\over {\partial \xi}}
+ {1\over {E^*-E_n(z)^*}}{{\partial E_n(z)^*}\over {\partial \xi}}
\right )\nonumber \\
=-\, {1\over {2i}}\sum_n \left (
{1\over {E-E_n(z)}}{{\partial E_n(z)}\over {\partial \varphi}}-
{1\over {E^*-E_n(z)^*}}{{\partial E_n(z)^*}\over {\partial \varphi}}
\right )
\label{ccf}
\end{eqnarray}
Let us denote with $N_+(E)$ and $N_-(E)$ the numbers of positive and negative 
exponents of $T(E)$. We have: \vskip 0.5truecm
\noindent
{\bf Proposition 5:} $N_+(E) = N_-(E) $\parno
Proof: for real $E$ we know that the exponents come in pairs $\pm \xi_a(E)$, 
because of proposition 1. Let us consider the case $Im E\neq 0$.\parno 
As a consequence of duality we derived that no eigenvalue of $T(E)$ is on the
unit circle, thus all exponents are non-zero: $N_- + N_+=2M$. Therefore, if 
we set $\xi =0$ in (\ref{cf}), the left term is  $N_-$. We now show that 
$N_-=M$ or, equivalently, that the integral in (\ref{cf}) vanishes for 
$\xi =0$. In the expression (\ref{ccf}) the eigenvalues $E_n(e^{i\varphi })$ 
are real periodic functions of $\varphi $ in $[0,2\pi]$, then 
\begin{eqnarray}
N_-(E) -M = \sum_{n=1}^{NM}\int_0^{2\pi} {{d\varphi}\over {2\pi}} 
{{dE_n}\over {d\varphi}} {{ {\rm Im}E}\over {({\rm Re}E-E_n )^2+
({\rm Im}E)^2}} =0 \qquad \bullet \nonumber
\end{eqnarray}        
\parno
As a function of $\varphi $ ($\xi $ is fixed) an eigenvalue 
$E_n(e^{\xi+i\varphi })$ makes a loop $\gamma_n$ in the complex $E$ plane. The 
loop $\gamma_n^*$ of $E_n^*$ is specular with respect respect of the real  
axis. Integration in $\varphi $ of (\ref{ccf}) yields Cauchy integrals 
\begin{eqnarray}
{\cal N}(\xi_a(E)<\xi)= M + {1\over 2}\sum_n \left (
\int_{\gamma_n}  {{dE^\prime}\over {2\pi i}}{1\over {E^\prime -E}} -
\int_{\gamma_n^*}  {{dE^\prime}\over {2\pi i}}{1\over {E^\prime-E^*}} \right )
\nonumber
\end{eqnarray}
The first integral is the winding number of the (oriented) trajectory 
$\gamma_n$ of $E_n(e^{\xi+i\varphi })$ 
around the value $E$. The second integral
is the winding number of $\gamma_n^*$ around $E^*$ and has opposite
sign because of opposite orientation.\parno   
We then obtain a nice geometric result:
\vskip 0.5truecm
\noindent 
{\bf Proposition 6:} ${\cal N}(\xi_a(E)<\xi )= M + {\cal W}(E) $\parno
{\sl {The number of exponents of $T(E)$ less than $\xi $ equals $M$ plus the 
total winding number of loops of eigenvalues $E_n (e^{\xi+i\varphi })$,
$-\pi<\varphi <\pi$, that encircle $E$}}
\vskip 0.5truecm\par
As I said, these formulae hold for a single general block tridiagonal
Hamiltonian matrix and its
transfer matrix. For a statistical ensemble of Hamiltonians one performs 
ensemble averages in place of a phase average, and deals with Lyapunov 
exponents. Souillard \cite{Cri} obtained the following formula for the 
positive Lyapunov exponents, which is the statistical analogue of 
(\ref{Soui1}):
\begin{eqnarray}
 {1\over M}\sum_a \gamma_a(E) = const. +  \int dE^\prime \rho (E^\prime ) 
\log |E-E^\prime|  
\end{eqnarray} 
$\rho (E)$ is the ensemble averaged spectral density of the Hermitian 
Hamiltonians. I am not aware of any statistical analogue of (\ref{cf}).
\vskip 0.5truecm

\noindent
{\bf {4. Bands, arcs and energy level motion.}}\parno
Spectral duality provides information on the positions of the 
eigenvalues of ${\cal H}(z)$ and their motion under variations of the 
boundary parameter $z$.\parno
It is useful to introduce the notion of {\bf discriminant}. If $z_a(E)$ is an 
eigenvalue of $T(E)$, the discriminant is the eigenvalue of $T(E)+T(E)^{-1}$
with same eigenvector:
\begin{eqnarray}
\Delta_a(E) = z_a(E)+{1\over {z_a(E)}} = 2\cosh \xi_a\cos\varphi_a + 2i
\sinh \xi_a \sin \varphi_a
\end{eqnarray}
Since the symplectic property implies that $1/z_a^*$ is an eigenvalue 
of $T(E^*)$, it is, in general: 
\begin{eqnarray}
\Delta_a (E^*)=\Delta_a(E)^*
\end{eqnarray} 
\vskip 0.5truecm

Let us begin with the simpler case where $H_i$, $L_i$ are real matrices.
Then $T(E^*)=T(E)^*$ and the eigenvalues of the transfer matrix come in 
pairs $z_a$ and $1/z_a$, $a=1\ldots M$.
Moreover, if $E$ is real the characteristic polynomial of $T(E)$ has real
coefficients and roots come also in pairs $z_a$, $z_a^*$.\parno
The spectral duality reads:
\begin{eqnarray}
 \prod_{a=1}^M \left [ \left (z_a+{1\over {z_a}}\right ) -\left ( z+{1\over z}
\right )\right ]  =
\det[L_1\ldots L_n]^{-1}\det [E-{\cal H}(z)] \label{41} 
\end{eqnarray}
Therefore the $M$ equations
\begin{eqnarray}
\Delta_a(E) = z+{1\over z} \label{16}
\end{eqnarray}
provide the $NM$ eigenvalues of ${\cal H}(z)$, which are naturally classified
in subsets with label $a$. We consider two cases: $|z|=1$ and $z$ real.\par
When $z=e^{i\varphi}$ all eigenvalues of ${\cal H}(e^{i\varphi})$ 
are real periodic functions of $-\pi\le \varphi<\pi$. 
Each equation $\Delta_a(E) =2\cos\varphi $ provides a number 
$n_a\ge 0$ of real solutions, and $n_1+\ldots +n_M=NM$.
This means that $y=\Delta_a(E) $, as a function of the real variable $E$, 
crosses $n_a$ times the strip $-2\le y\le 2$ parallel to the $E$ axis.
No extrema are allowed in the strip, for all branches must cut $n_a$ times 
the lines 
$y=2\cos\varphi $, to ensure that ${\cal H}(e^{i\varphi})$ 
has $NM$ real eigenvalues for all $\varphi$.  All branches of the
functions $\Delta_a(E)$ cross their bands in the eigenvalues of 
${\cal H}(i)$. See Fig. 1.\parno
By changing $\varphi $ the eigenvalues span bands in the real axis 
$$     E=E_{a,j}(\varphi) \qquad a=1\ldots M,\,\,  j=1\ldots n_a $$  
with extrema given by the eigenvalues of the periodic ($\varphi=0$) and
antiperiodic ($\varphi=\pi$) Hamiltonians. The velociy of level motion is
\begin{eqnarray}
{{\partial E}\over {\partial\varphi}} = -{{2\sin\varphi}\over 
{\Delta_a^\prime (E)}}
\end{eqnarray}
In the turning points the velocity vanishes. The second derivative is
known as the curvature \cite{Cas,Zycz,Gua}. This dynamics is invariant under 
the  ``time-reversal'' operation  $\varphi\to -\varphi $, corresponding 
to the transposition of the Hamiltonian matrix.\parno
Bands with same label $a$ may at most share an extremum, while bands 
related to different $a$ may overlap.\parno
When a branch of $\Delta_a$ and a branch of $\Delta_{a^\prime}$
cross inside the strip in a point $(E,2\cos\varphi )$, there is a crossing
of two eigenvalues of ${\cal H}(e^{i\varphi})$ (and a collision of two pairs
of eigenvalues of $T(E)$ in the unit circle). This is a highly non-generic 
occurrence for one-parameter Hermitian matrices.\parno
Energy bands of Hermitian periodic tridiagonal matrices ($M=1$) 
were studied in \cite{Koro}\parno
\\
\epsfig{file=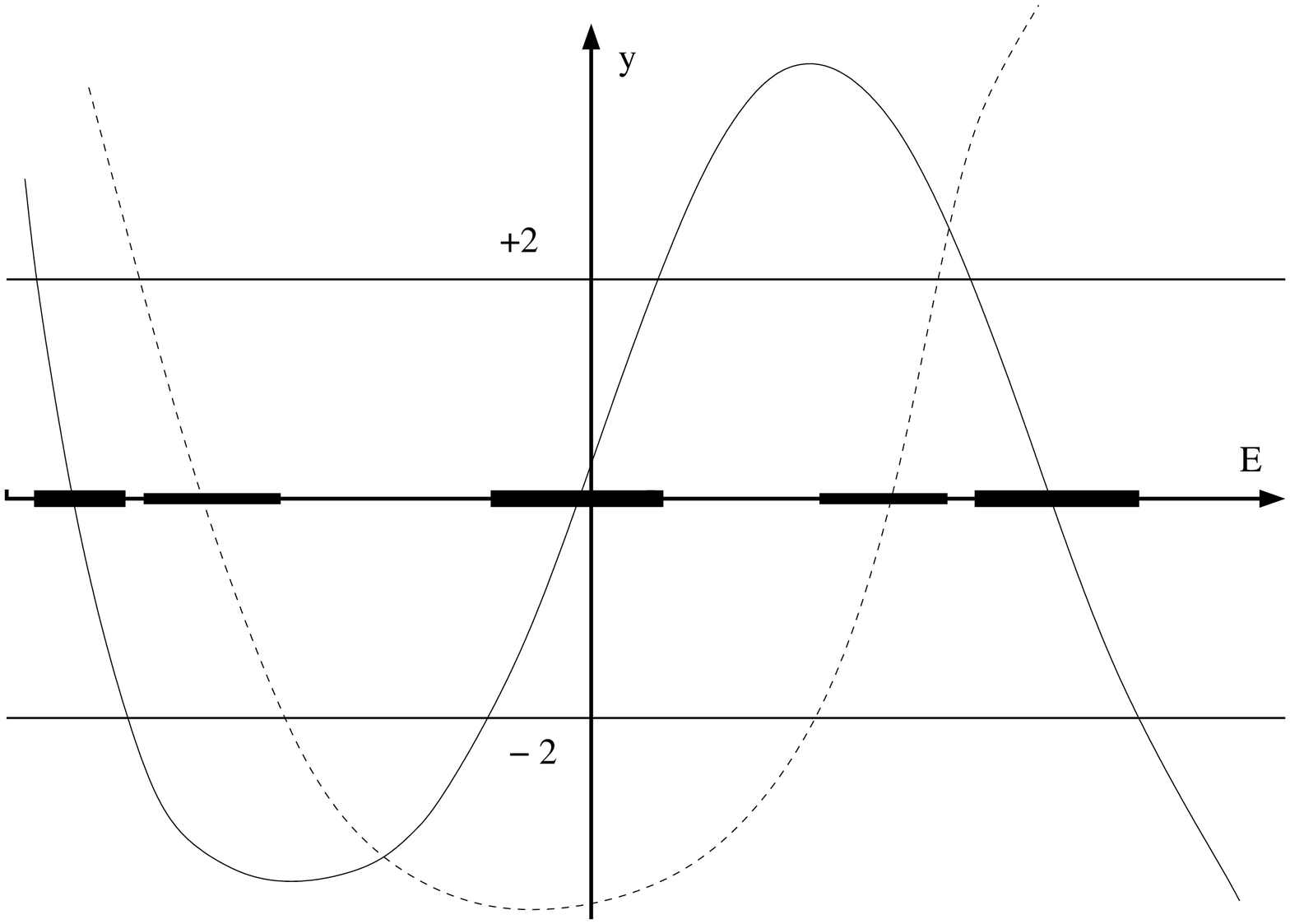,height=6cm}\\ 
{\sl Fig.1 Energy bands of a pentadiagonal matrix (M=2), solution to
$\Delta_a(E)=2\cos\varphi$, $a=1,2$, $-\pi\le \varphi <\pi $.}\\
\\
\par
When $z=e^{Ng}$, with $g\ge 0$, the matrix ${\cal H}(z)$ is real. 
The equations $\Delta_a(E)=2\cosh (Ng)$ provide all $NM$ eigenvalues in 
subsets labelled by $a$. They imply the equation $\xi_a(E)=\pm Ng$. 
For finite $g$ and large $N$, given that the exponents grow linearly in $N$ 
with a coefficient that defines the Lyapunov exponent, the eigenvalues of the 
non-Hermitian matrix ${\cal H}(z)$ distribute along $M$ level lines 
\begin{eqnarray}
\gamma_a(E)=g
\end{eqnarray}
The distribution along arcs is shown in Fig. 2, for large $z$.
For small $z$ the pattern of the eigenvalue distribution  
is complicate and intertwined.\parno 
For $g$ small enough, in continuity with the description given
for the periodic case, the eigenvalues still belong to the real axis, outside
their periodicity bands. By increasing $g$, eigenvalues with the same 
label $a$ pairwise approach until a pair condenses. Correspondingly, 
an extremum is reached for the function $\Delta_a(E)$ and the pair of 
eigenvalues acquires opposite imaginary parts. 
There is a critical value $g_a$ of $g$ for this to happen for each label $a$: 
\begin{eqnarray}
 \Delta_a^\prime (E_a)=0 \quad \Delta_a(E_a) = 2\cosh (Ng_a) 
\end{eqnarray}
\epsfig{file=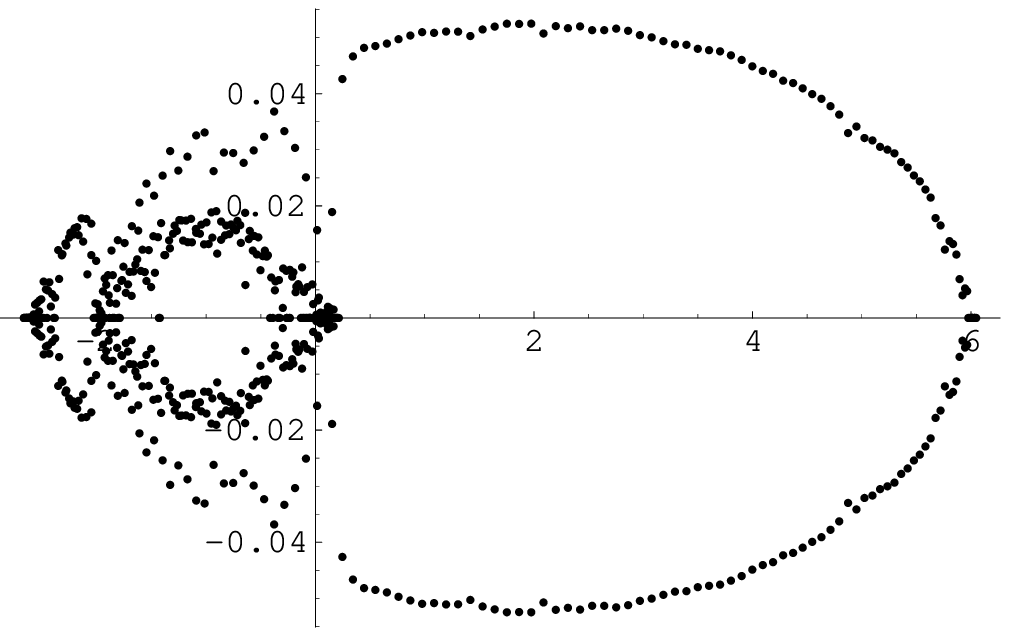,height=3.6cm}\epsfig{file=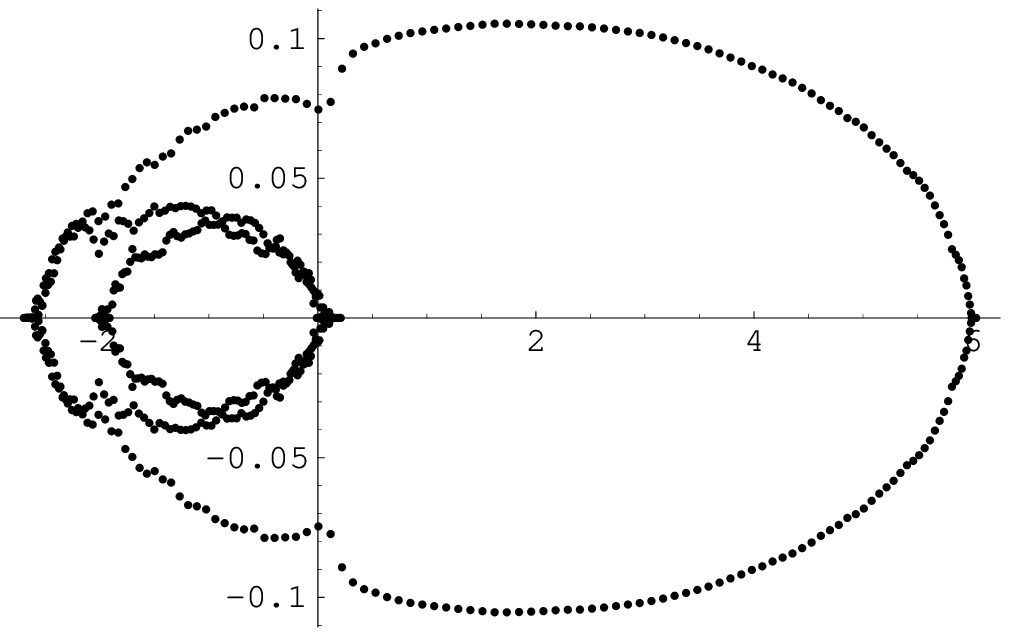,height=3.6cm}\\ 
{\sl Fig.2  The 600 eigenvalues of an eptadiagonal matrix (M=3, N=200),
with diagonal uniform disorder $|a_i|<0.5$, unit off-diagonal elements, and 
$z=20$ (left), $z=390$ (right).}\\
\par
Finally, let us mention the case where $H_i=H_i^\dagger $ and $L_i$ are not 
real matrices. Another spectral identity is needed, which follows 
from  spectral duality and the symplectic property \cite{Mol1}. It holds 
for any value of $z$ or $E$ in the complex plane:
\begin{eqnarray}
 \det [ T(E)+T(E)^{-1} -(z+{1\over z})]= |\det L_1\ldots L_n|^{-2}
\det [E-{\cal H}(z)]\det [E-{\cal H}({1\over z})] 
\end{eqnarray}
For $z=e^{i\varphi }$ the right side is zero in $2NM$ real solutions, 
$E_i(\varphi )$ and $E_i(-\varphi )$, $i=1\ldots NM$, that 
span the same $NM$ bands as $\varphi $ varies in $-\pi, \pi$. They
are degenerate in $\varphi =0$ ($y=2$) and $\varphi =\pi$ ($y=-2$).\parno
The strip $-2\le y\le 2$ is thus crossed by $2NM$ branches $y=\Delta_a(E)$ 
which join pairwise at the boundaries $y=\pm 2$ of the strip. Each pair,
when intersected with the line $y=2\cos\varphi $, determines the
same band which the eigenvalue $E(\varphi )$ covers with different speeds 
in the two directions.\par
For $z=\pm e^{Ng}$ we have again condensation of pairs of eigenvalues,
which no longer become complex conjugate pairs. For large $N$ the 
eigenvalues move into $M$ level curves $\gamma_a(E)=g$ (the case
$g<0$ leads to a different set of curves since in this case Lyapunov
exponents need not be opposite pairs). 
\vskip 0.5truecm

\noindent
{\bf Conclusion}\par
The spectral duality is a simple identity that links the eigenvalues of a 
matrix with block-tridiagonal structure to those of the corresponding 
transfer matrix. I have deduced some analytic properties for the
exponents (sum of exponents, $N_+=N_-$, counting function). They involve
a phase average on eigenvalues of the block-matrix.\parno
The large $N$ stability of exponents allows to describe the
distribution of complex eigenvalues of the block matrix along arcs.
Discriminants classify real eigenvalues (periodic case) in bands, and govern
their motion and collisions.\parno 
These exact properties are hoped to allow a more analytic approach to the 
difficult study of Lyapunov spectra of transfer matrices, that are 
{\sl derived} from an ensemble of random Hamiltonians. They also extend 
to block matrices some results which were known for purely tridiagonal 
matrices.
\vskip 1truecm

\vfill
\end{document}